\documentclass[aps,prl,twocolumn]{revtex4-1}

\usepackage{times}

\usepackage{color}

\usepackage{multirow}

\newcommand{\unit}[1]{\ensuremath{\;\mathrm{#1}}}

\newcommand{\ket}[1]{|#1\rangle}

\newcommand{\SH}{\mathbf{H}}
\newcommand{\SV}{\mathbf{V}}
\newcommand{\SD}{\mathbf{D}}
\newcommand{\SA}{\mathbf{A}}
\newcommand{\SR}{\mathbf{R}}
\newcommand{\SL}{\mathbf{L}}

\usepackage{graphicx}
\usepackage{hyperref} % has to be last command in preamble to work right
\hypersetup{bookmarks=true,colorlinks=true,linkcolor=blue,citecolor=green}

\begin{document}

% Include your paper's title here
\title{Experimental Three-Particle Quantum Nonlocality under Strict Locality Conditions}

%\author{C. Erven$^{1,6}$, E. Meyer-Scott$^{1}$, K. Fisher$^{1}$, J. Lavoie$^{1}$, B. L. Higgins$^{1}$, Z. Yan$^{1,2}$, C. J. Pugh$^{1}$, J.-P. Bourgoin$^{1}$, R. Prevedel$^{1,3}$, L. K. Shalm$^{1,4}$, L. Richards$^{1}$, N. Gigov$^{1}$, R. Laflamme$^{1}$, G. Weihs$^{5}$, T. Jennewein$^{1}$ \& K. J. Resch$^{1\dag}$}

%\author{NN}
\author{C. Erven$^{1,6}$}
\email[]{chris.erven@bristol.ac.uk}

\author{E. Meyer-Scott$^{1}$}
\author{K. Fisher$^{1}$}
\author{J. Lavoie$^{1}$}
\author{B. L. Higgins$^{1}$}

\author{Z. Yan$^{1,2}$}

\author{C. J. Pugh$^{1}$}
\author{J.-P. Bourgoin$^{1}$}

\author{R. Prevedel$^{1,3}$}
\author{L. K. Shalm$^{1,4}$}

\author{L. Richards$^{1}$}
\author{N. Gigov$^{1}$}

\author{R. Laflamme$^{1}$}
\author{G. Weihs$^{5}$}

\author{T. Jennewein$^{1}$}
\author{K. J. Resch$^{1}$}
\email[]{kresch@uwaterloo.ca}

\affiliation{$^{1}$Institute for Quantum Computing and Department of Physics \& Astronomy, University of Waterloo, Waterloo, ON, N2L 3G1, Canada}
\affiliation{$^{2}$Centre for Ultrahigh Bandwidth Devices for Optical Systems (CUDOS) \& MQ Photonics Research Centre, Department of Physics \& Astronomy, Macquarie University, Sydney, NSW 2109, Australia}
\affiliation{$^3$Research Institute of Molecular Pathology and Max F. Perutz Laboratories GmbH, Dr. Bohr-Gasse 7-9, 1030, Vienna, Austria}
\affiliation{$^4$National Institute of Standards and Technology, Boulder, CO 80305, USA}
\affiliation{$^5$Institut f\"ur Experimentalphysik, Universit\"at Innsbruck, Technikerstrasse 25, 6020 Innsbruck, Austria}
\affiliation{$^6$Centre for Quantum Photonics, H. H. Wills Physics Laboratory \& Department of Electrical and Electronic Engineering, University of Bristol, Merchant Venturers Building, Woodland Road, Bristol, BS8 1UB, UK}

% Include the date command, but leave its argument blank.
%\date{}
\date{\today}

\maketitle

%===============================================================================
% Body
%===============================================================================

\textbf{
Quantum correlations are critical to our understanding of nature, with far-reaching technological\cite{BB84,SBCDLP09,LJLNMO10,NC00} and fundamental impact. These often manifest as violations of Bell's inequalities\cite{Bel64,CHSH69,FC72,AGR82}, bounds derived from the assumptions of locality and realism, concepts integral to classical physics. Many tests of Bell's inequalities have studied pairs of correlated particles; however, the immense interest in multi-particle quantum correlations is driving the experimental frontier to test systems beyond just pairs. All experimental violations of Bell's inequalities to date require supplementary assumptions, opening the results to one or more loopholes, the closing of which is one of the most important challenges in quantum science. Individual loopholes have been closed in experiments with pairs of particles\cite{RKMSIMW01,GMRWKBLCGNUZ13,CMACGLMSZNBLGK13,ADR82,WJSWZ98,SUKRMHRFLJZ10} and a very recent result closed the detection loophole in a six ion experiment\cite{BBSNHMGB13}. No experiment thus far has closed the locality loopholes with three or more particles. Here, we distribute three-photon Greenberger-Horne-Zeilinger entangled states\cite{GHZ89} using optical fibre and free-space links to independent measurement stations. The measured correlations constitute a test of Mermin's inequality\cite{Mer90} while closing both the locality and related freedom-of-choice loopholes due to our experimental configuration and timing. We measured a Mermin parameter of $2.77 \pm 0.08$, violating the inequality bound of $2$ by over $9$ standard deviations, with minimum tolerances for the locality and freedom-of-choice loopholes of $264 \pm 28$\unit{ns} and $304 \pm 25$\unit{ns}, respectively.  These results represent a significant advance towards definitive tests of the foundations of quantum mechanics\cite{PCLWZZ12} and practical multi-party quantum communications protocols\cite{HBB99}.
}

%\section*{Introduction}

In his breakthrough work, John Bell\cite{Bel64} derived upper bounds on the strength of correlations exhibited by local hidden variable (LHV) theories, very general models of nature in which measurement outcomes in one region of space are independent of the events in other space-like separated regions. Quantum mechanical correlations can violate these bounds. Greenberger, Horne, and Zeilinger (GHZ) extended Bell's argument to the scenario with three-particle entangled states, and showed they could manifest violations of local realism in a fundamentally different way\cite{GHZ89,GHSZ90}. The GHZ argument was converted into the form of an inequality by Mermin\cite{Mer90} which we experimentally tested.

An ideal Bell inequality experiment requires separating two or more particles by a large distance and making high-efficiency local measurements on those particles using randomly chosen settings and comparing these results\cite{Bel81,Per70}. The first Bell inequality tests were carried out using two-photon cascades in atomic systems\cite{FC72,AGR82}. Mermin's inequality was violated using three-photon entanglement from a parametric down-conversion source\cite{PBDWZ00}. However, these and the many other Bell experiments that followed are subject to one or more loopholes that could, in principle, be exploited to yield a violation even though nature is in fact describable by an LHV model. The \emph{detection loophole} concerns LHV models which take advantage of low detection efficiency\cite{Per70}; it has been closed for two-particle Bell inequalities in ion trap and photon experiments\cite{RKMSIMW01,GMRWKBLCGNUZ13, CMACGLMSZNBLGK13} as well as a six ion experiment\cite{BBSNHMGB13}. Those experiments that do not close this loophole must appeal to the \emph{fair-sampling assumption}: properties of the subset of particles that are measured are representative of all the particles. The \emph{locality loophole} (L) exploits configurations where the choice of measurement settings and measurement outcomes at the distant locations are not causally disconnected, i.e., not outside each others' forward and backward light cones. This loophole has been closed for two-particle Bell inequalities in photon experiments\cite{ADR82,WJSWZ98,SUKRMHRFLJZ10}. The \emph{freedom-of-choice loophole} (FoC) exploits arrangements where the choice of measurement settings is not causally disconnected from the source. This loophole was recently identified and closed using photons\cite{SUKRMHRFLJZ10}. A loophole-free Bell inequality test has yet to be performed.

No attempts have been made to close locality loopholes in Bell experiments involving three or more particles\cite{PBDWZ00,ZYCZZP03,LKR09}. The primary reason for this is source brightness. While entangled photon pairs have been generated and detected at rates in excess of 1\unit{MHz}\cite{AJK05,FHPJZ07} entangled photon triplets have only been observed at rates on the order of Hz\cite{HHFRRJ10,LHB01}, necessitating long measurement times. In addition, further experimental challenges include high sensitivity to loss, causality relations requiring a complex experimental setup, and demanding stability.

% Figure captions should be 100 words or less.
\begin{figure*}[htbp]
    \centering
    \includegraphics[width=1.85\columnwidth]{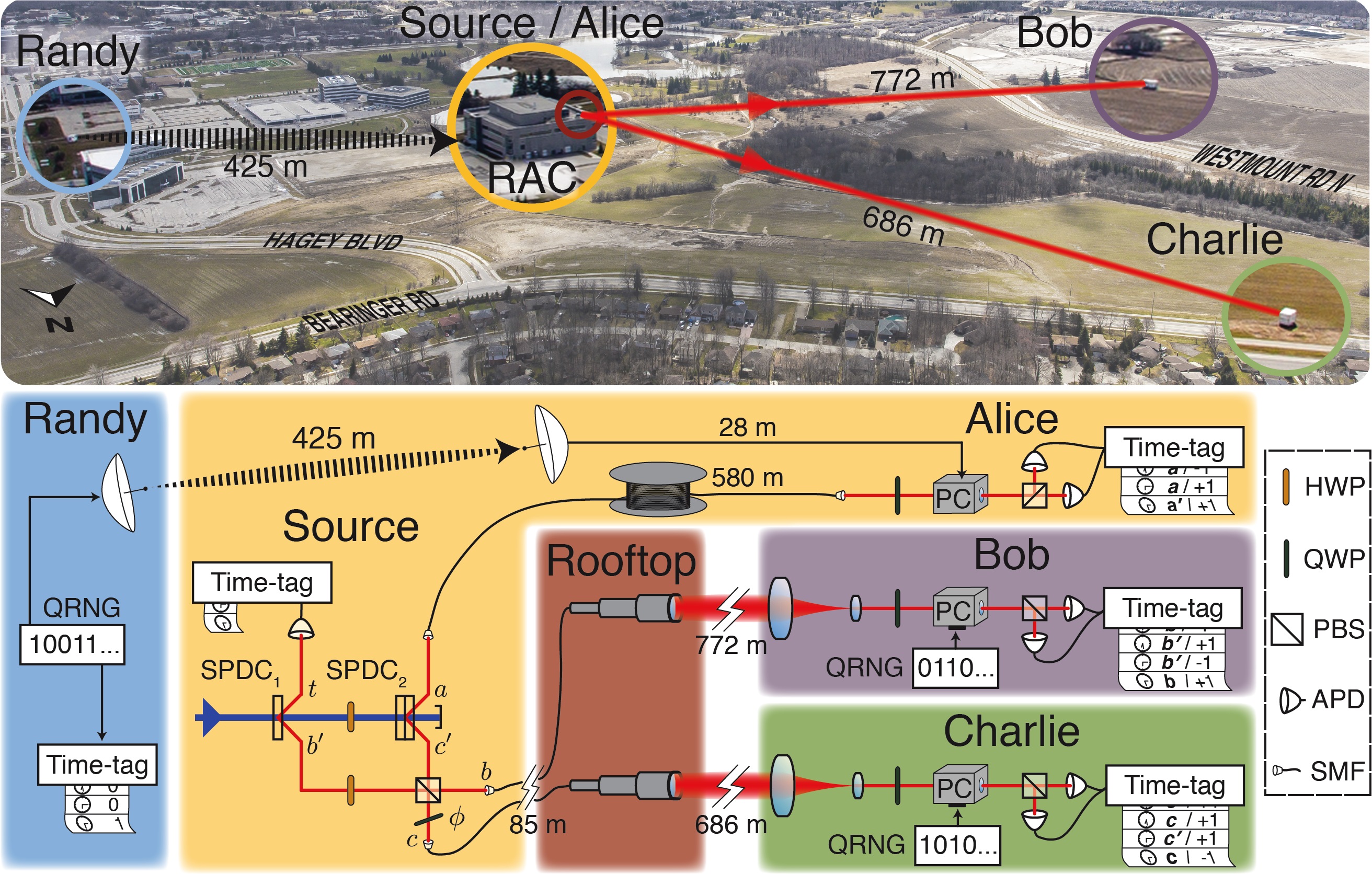}
    \caption{Experimental setup. The down-conversion (SPDC) source of triggered three-photon GHZ states and Alice were located in the RAC building. Two of the entangled photons were sent to the roof through optical fibers and transmitted to trailers, Bob and Charlie, through free-space optical links. Bob and Charlie were 801\unit{m} and 721\unit{m} away from the source, respectively; the optical links used to send their photons were 772\unit{m} and 686\unit{m} long, respectively. Bob and Charlie measured their photon polarizations in one of two bases, determined by co-located QRNGs and fast Pockels cells, and recorded measurement values using time-tagging electronics. A third trailer, Randy 446\unit{m} from the source, contained a QRNG which sent random bits to Alice over a 425\unit{m} RF link determining the setting of Alice's Pockels cell. Randy contained time-tagging electronics to record the output of the QRNG for comparison with Alice's record. Alice's photon was delayed in a fiber spool before its polarization was measured and the result recorded using time-tagging electronics. Half-wave plate (HWP); quarter-wave plate (QWP); polarizing beam-splitter (PBS); avalanche photo-diode (APD); single-mode fibre (SMF).}
    \label{fig.MapAndSchematic}
\end{figure*}

Here we have overcome these challenges and report the experimental violation of the three-particle Mermin's inequality closing both the locality and freedom-of-choice loopholes, having to make only the fair-sampling assumption. We created three-photon polarization-entangled states and sent two of the photons to different distant locations. The correlations were extracted from measurements of each photon polarization using randomly chosen settings. Our comprehensive analysis of the space-time arrangement of the experiment establishes independence of the important events.

Mermin considered a scenario where three particles were measured at stations which we call \emph{Alice}, \emph{Bob}, and \emph{Charlie}. Each particle was measured by a device with two settings, e.g. $\mathbf{a}$ and $\mathbf{a'}$ for the device at Alice, and two outcomes, $+1$ and $-1$. The correlation, $E$, in the outcomes for the settings $\mathbf{a}$, $\mathbf{b}$, and $\mathbf{c}$ is defined as $E(\mathbf{a},\mathbf{b},\mathbf{c})= P^{+} - P^{-}$, where $P^{+}$ ($P^{-}$) is the probability that the product of the outcomes is $+1$ ($-1$). LHV theories must obey Mermin's inequality\cite{Mer90},
\begin{eqnarray}\label{eq.MerminInequality}
  M & = & |E(\mathbf{a},\mathbf{b},\mathbf{c}) - E(\mathbf{a},\mathbf{b'},\mathbf{c'}) \nonumber \\ 
   & & - E(\mathbf{a'},\mathbf{b},\mathbf{c'}) - E(\mathbf{a'},\mathbf{b'},\mathbf{c)}| \leq 2, 
\end{eqnarray}
where $M$ is the \emph{Mermin parameter}.

Mermin's inequality can be maximally violated using three-particle GHZ states encoded, for example, in photon polarization,
\begin{equation}\label{eq.GHZState}
  \ket{\mathrm{GHZ}} = \frac{1}{\sqrt{2}}\left(\ket{\SH \SH \SH} - i \ket{\SV \SV \SV}\right),
\end{equation}
where $\ket{\SH}$ and $\ket{\SV}$ represent horizontal and vertical polarizations respectively. We define the diagonal/antidiagonal ($D/A$) states as $\ket{\SD} = \frac{1}{\sqrt{2}} (\ket{\SH} + \ket{\SV})$ and $\ket{\SA} = \frac{1}{\sqrt{2}} (\ket{\SH} - \ket{\SV})$, and right/left ($R/L$) states as $\ket{\SR} = \frac{1}{\sqrt{2}} (\ket{\SH} +i \ket{\SV})$ and $\ket{\SL} = \frac{1}{\sqrt{2}} (\ket{\SH} - i\ket{\SV})$. When we measure $\ket{\SD}$ or $\ket{\SR}$, we assign the outcome $+1$; when we measure $\ket{\SA}$ or $\ket{\SL}$, we assign $-1$.  If we measure in the $R/L$ basis for settings $\mathbf{a}$, $\mathbf{b}$, and $\mathbf{c}$ and in the $D/A$ basis for setting $\mathbf{a'}$, $\mathbf{b'}$, and $\mathbf{c'}$ then quantum theory predicts a Mermin parameter of 4, the maximum violation of the inequality.

%\section*{Experiment}

Our experimental configuration is shown in Fig.~\ref{fig.MapAndSchematic}. The entangled photon source was located in a laboratory in the Research Advancement Centre (RAC) building. The source produced three-photon entangled states and a fourth \emph{trigger} photon which was detected locally. The four-photon coincidence rate was 39\unit{Hz}, measured at the source. Photons from the entangled state were sent to three measurement stations, Alice, Bob, and Charlie. Photons were distributed to Bob and Charlie, located in trailers approximately 801\unit{m} west and 721\unit{m} northwest of the source, via free-space links. Alice was co-located with the source, and her photon was delayed in an optical fibre. The measurement basis choice for Alice was made by a fourth party, \emph{Randy}, located in a trailer 446\unit{m} east of the source. Each receiver measured the polarization in one of two bases, chosen by a fast quantum random number generator (QRNG)\cite{JAWWZ00}. All single-photon detection events and Randy's QRNG results were recorded using time-tagging electronics. For more details refer to the Supplementary Information.

To test Mermin's inequality, we recorded time-tags for 1\unit{hr} 19\unit{min}, while the receivers measured either in the $R/L$ or $D/A$ bases. Over this time the free-space link efficiencies averaged 33\% and 32\% to Bob and Charlie, respectively; the long fibre efficiency at Alice was 14\%. Once the experiment was completed, we extracted all four-fold events using a 3\unit{ns} coincidence window. This yielded 2,472 four-fold coincidence events, corresponding to an average rate of 0.5\unit{Hz}. The raw counts for each polarization setting are given in the Supplementary Information. Fig.~\ref{fig.ExpectationsMin} shows the measured correlations which constitutes a Mermin parameter of $2.77 \pm 0.08$, where the uncertainty is based on Poissonian count statistics. These results significantly violate Mermin's inequality by over $9\sigma$. Furthermore, for the measurement settings which we employed, our violation of Mermin's inequality shows that our state was genuine tripartite entangled\cite{GT09}.

\begin{figure}[htbp]
    \centering
    \includegraphics[width=0.95\columnwidth]{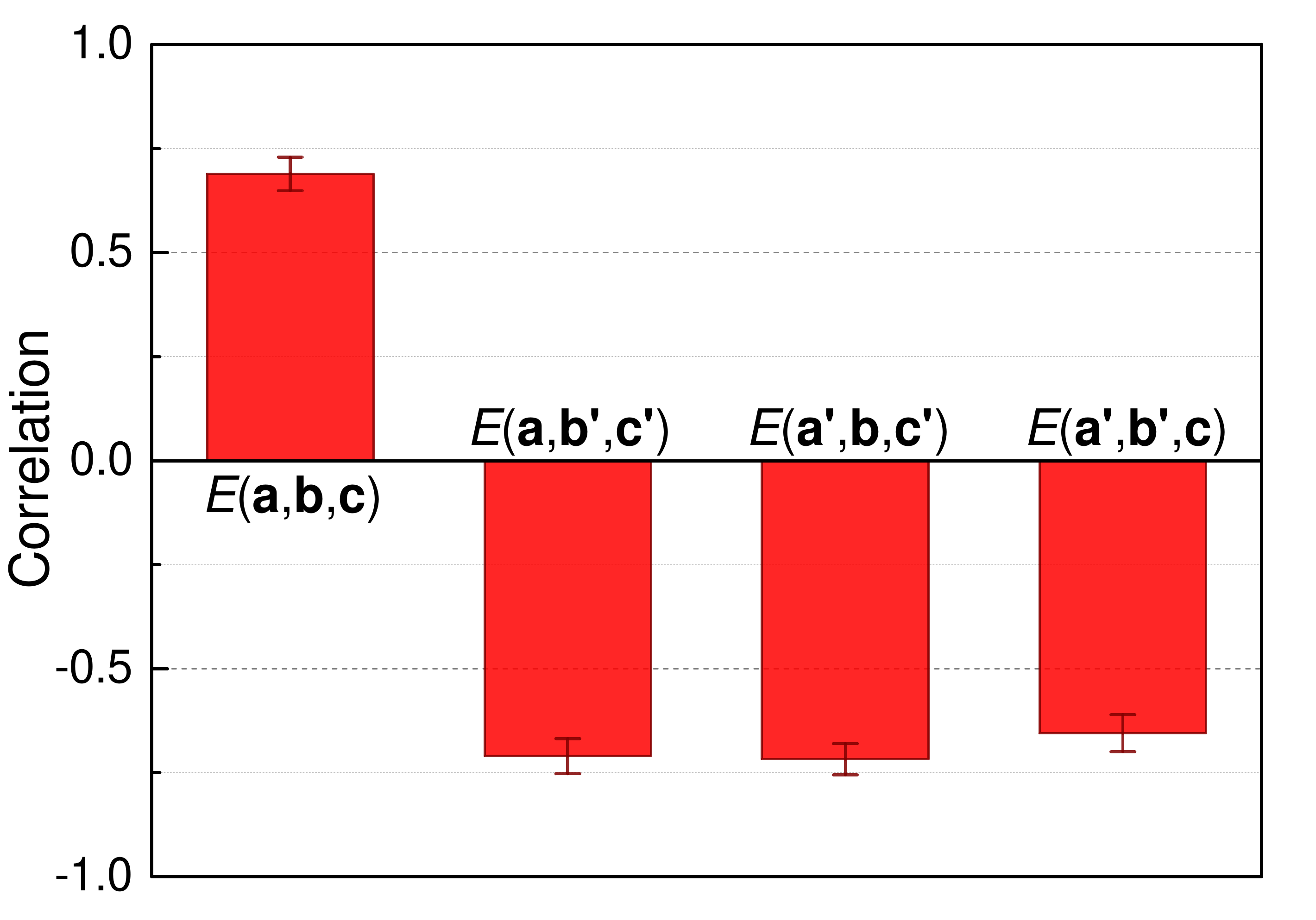}
    \caption{Experimentally measured three-photon polarization correlations. The analyzers were set to the \emph{R/L} basis for settings $\mathbf{a}$, $\mathbf{b}$, and $\mathbf{c}$ and the \emph{D/A} basis for settings $\mathbf{a'}$, $\mathbf{b'}$, and $\mathbf{c'}$. During a 1\unit{hr} 19\unit{min} experiment, we measured 2,472 four-fold coincidence events of which 1,232 were used to extract correlations for a test of Mermin's inequality. The measured correlations were $E(\mathbf{a},\mathbf{b},\mathbf{c}) = 0.689 \pm 0.040$, $E(\mathbf{a},\mathbf{b'},\mathbf{c'}) = -0.710 \pm 0.042$, $E(\mathbf{a'},\mathbf{b},\mathbf{c'}) = -0.718 \pm 0.038$, and $E(\mathbf{a'},\mathbf{b'},\mathbf{c}) = -0.655 \pm 0.044$ yielding a Mermin parameter of $2.77 \pm 0.08$ violating the local hidden variable bound of 2 by over 9$\sigma$. Error bars represent one standard deviation based on Poisson statistics.}
    \label{fig.ExpectationsMin}
\end{figure}

We performed additional tests of Mermin's inequality with different phase settings in the GHZ state; in all cases we measured a significant violation and the correlations depended on the phase as expected. The results are described in the Supplementary Information.

%\section*{Space-time Analysis}

The timing and layout of the experiment are both critical for closing the locality and freedom-of-choice loopholes. The space-time analysis of our setup includes: the locations of the components, measured delays, delays inferred from distance measurements, autocorrelation times of the QRNGs, and the asynchronous QRNG sampling. This analysis is summarized in Fig.~\ref{fig.Spacetime} and fully detailed in the Supplementary Information. In Fig.~\ref{fig.Spacetime}~(a) the distances and angles from the Source/Alice to Bob, Charlie, and Randy are shown. Note that Fig.~\ref{fig.MapAndSchematic} shows the free-space optical link distances whereas Fig.~\ref{fig.Spacetime} shows the straight-line distances between the source and stations.

\begin{figure*}[htbp]
    \centering
    \includegraphics[width=1.85\columnwidth]{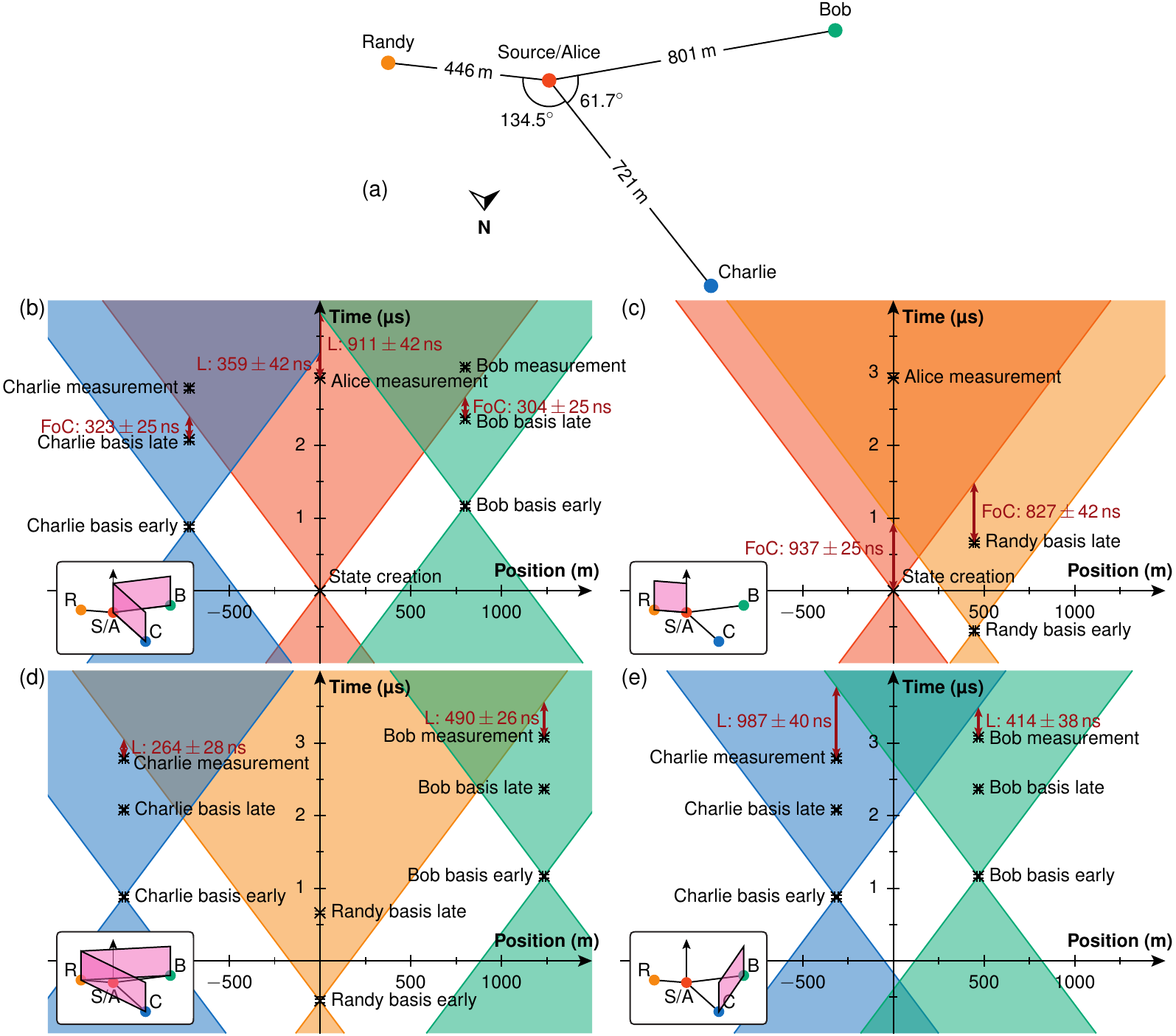}
    \caption{Space-time analysis of the experiment. (a) Simplified layout of the experiment showing the Euclidean distances and angles between locations. (b)--(e) Six 2D space-time diagrams, one for each pair of locations, which fully describe the relationship between important events in the experiment in the lab frame, namely the entangled photon creation event labelled \emph{State creation}, choices of measurement bases, and the measurements themselves. The insets show the relevant locations for each panel. The earliest and latest times at which Bob's measurement basis could have been selected are labelled \emph{Bob basis early} and \emph{Bob basis late}, respectively. Labelling at Randy and Charlie follow this convention. The event corresponding to the measurement of a photon from the source at Bob is labelled \emph{Bob measurement}, similarly for Alice and Charlie. The light-cones for the important events are shown as diagonal lines and shaded regions for Source/Alice (red), Bob (green), Charlie (blue), and Randy (orange). Red vertical arrows indicate tolerances for the freedom-of-choice (FoC) and locality (L) conditions. The minimum tolerance for the FoC is $304 \pm 25$\unit{ns} as seen in panel (b) and the minimum for the locality condition is $264 \pm 28$\unit{ns} as seen in panel (d), indicating that both conditions are met by a significant margin.}
    \label{fig.Spacetime}
\end{figure*}

The space-time description of the experiment is inherently 3+1-dimensional, however we can extract all relevant information from six 2-dimensional slices, one for each pair of locations, shown in Fig.~\ref{fig.Spacetime}~(b)--(e). All times and positions are given in the laboratory frame of reference. For compactness, Fig.~\ref{fig.Spacetime}~(b) and (d) each depict two 2D slices, however, since the locations are non-collinear the two halves of the diagram (positive and negative position) should be interpreted independently.

For example, in Fig.~\ref{fig.Spacetime}~(b), the GHZ state is produced at the origin and labelled \emph{State creation}. The red diagonal lines emanating from the origin define the past and future light cones of the source production event. Those events on or above the future light cone could be causally influenced by the source according to special relativity, while those events on or below the past light cone could causally influence the source. Those events outside the light cones of the source are independent and cannot influence nor be influenced by the source.

Now consider the right half of Fig.~\ref{fig.Spacetime}~(b), describing events relevant to the Source/Alice and Bob. Bob's events take place at his position 801\unit{m} from the source. The photon arrives at Bob's Pockels cell at time $t = 3032$\unit{ns}. We consider the measurement to be complete when the photon detection event is stored in the memory of the time-tagging electronics; adding the optical and electronic delays, the measurement at Bob occurs at $t = 3079 \pm 24$\unit{ns} and is labelled \emph{Bob measurement}.

We consider the random number generated in the QRNG when a photon from a light-emitting diode (LED) hits the beamsplitter inside the device. The latest time that the QRNG output could determine the measurement setting for his received photon is 665\unit{ns} before the photon passes through the PC (i.e., at $t = 2367$\unit{ns}); this is labelled \emph{Bob basis late}. This event occurs $304 \pm 25$\unit{ns} outside the source light cone, thus satisfying the FoC condition in this case. The earliest time at which the QRNG determines the setting is 1200\unit{ns} earlier at $t = 1167$\unit{ns}, which includes 1000\unit{ns} for asynchronous sampling of the QRNGs, and 200\unit{ns} for the QRNG autocorrelation, and is labelled \emph{Bob basis early}. Alice is co-located with the source at position 0\unit{m} and her measurement labelled \emph{Alice measurement} occurs at $t = 2926$\unit{ns}. This event is $911 \pm 42$\unit{ns} outside the light cone (green line) of the earliest basis setting at Bob, satisfying the locality condition here.

We summarize results for the other five slices shown in Fig.~\ref{fig.Spacetime}~(b)--(e), with labels defined using the same convention. In all cases, the basis choice is made outside the light cone from the source creation event by at least $304 \pm 25$\unit{ns} ensuring the FoC loophole is closed. The measurement at any location is completed outside the light cone of the earliest possible basis choice at each other location by at least $264 \pm 28$\unit{ns} ensuring the locality loophole is closed.

To test if the observed correlations depend on fast random switching, we performed an additional experiment comparing the case of the measurement bases selection by QRNGs to the case where they were deterministically selected with 500\unit{kHz} periodic signals from function generators. Mermin's inequality was violated in both cases indicating no significant difference between deterministic and random switching. These results are in the Supplementary Information.

%\section*{Conclusions}

Our results suggest several interesting avenues for future research, such as extending the setup to close the locality loopholes for Mermin inequalities with larger numbers of particles, testing more general non-local hidden-variable models\cite{Sve87}, and studying the improvements in source technology and link efficiency required to perform a loophole-free Mermin inequality. Our setup yielded a Mermin parameter only slightly reduced from that estimated from quantum state tomography measured directly at the source, demonstrating that multi-photon entanglement was distributed with high fidelity. Our experiment is thus an exciting new platform for multi-party quantum communications protocols\cite{HBB99}.

\section*{Supplementary Information: Experimental Three-Particle Quantum Nonlocality under Strict Locality Conditions}

\section*{Methods}

We produced two sets of photon pairs with wavelengths of 790\unit{nm} using spontaneous parametric down-conversion (SPDC) from two consecutive sources pumped by a pulsed laser. Combining one photon from each pair in a linear optics setup, we produced three-photon polarization-entangled GHZ states in modes $a$, $b$, and $c$ with a trigger photon in mode $t$, and coupled each photon into single-mode fiber. Locally, we measured singles rates of 576\unit{kHz}, 386\unit{kHz}, 417\unit{kHz}, and 476\unit{kHz} for modes \emph{a}, \emph{b}, \emph{c}, and \emph{t} respectively, and a four-fold coincidence rate of 39\unit{Hz} by connecting the short fibers directly to the avalanche photodiode (APD) detectors. From quantum state tomography \cite{JFH03}, our source fidelity was $(82.9 \pm 0.3)\%$ with the target GHZ state. Based on this reconstruction, the maximum Mermin parameter was $M = 3.08$ for optimal measurements.

The photons from the source were distributed to the measurement stations. The trigger photon was connected via a 2\unit{m} optical fiber directly to a detector. The photon in mode $a$ was connected to a 580\unit{m} fiber spool followed by a receiver located on the source optical table. The photons in modes $b$ and $c$ were sent through two 85\unit{m} fibers to separate sending telescopes inside a dome on the RAC rooftop and transmitted over 772\unit{m} and 686\unit{m} free-space links to receiver telescopes at Bob and Charlie, respectively. Each receiver at Alice, Bob, and Charlie measured the polarization of the photons in one of two bases. Each basis choice was determined by a quantum random number generator (QRNG) \cite{JAWWZ00}. In each QRNG, photons from a light emitting diode (LED) light were detected by one of two photon counters after a beamsplitter, resulting in random bits then sampled at a rate of 1\unit{MHz}; these bits controlled a Pockels cell (PC) thereby determining the polarization measurement. At Bob and Charlie, the QRNGs were co-located with the receivers. In Alice's case, basis choices were determined by the QRNG at Randy, sent to Alice via a 425\unit{m} radio-frequency (RF) link to a receiver outside the RAC building. All detection events and QRNG outputs were recorded at each station using time-tagging electronics with a precision of 156.25\unit{ps}, referenced using a 10\unit{MHz} clock from GPS units.

The receivers were initially set to measure either the $H/V$ or $R/L$ bases depending on the QRNG setting. In a night-time experiment, we aligned the free-space links using visible and infrared laser diodes. We ensured horizontal and vertical polarizations were faithfully transmitted by adjusting polarization controllers on local count rates at each station. In the ideal case, this ensured the receivers measured a state of the form of Eq.~2, up to a relative phase. Setting this phase was more difficult as it is a global property of the GHZ state and cannot be fixed by observing local counts. Therefore we measured four-photon coincidences between trigger, Alice, Bob, and Charlie in real time. All time-tags were sent to the computer at RAC over the Internet. Incorporating the delays in the experiment, custom software found four-photon detection events for each combination of polarizations and calculated the polarization correlations. The relative phase in the GHZ state was set by tilting a quarter-wave plate (QWP) in mode $c$ to maximize the four-fold rates for triggered $\ket{\SR\SR\SR}$, $\ket{\SR\SL\SL}$, $\ket{\SL\SR\SL}$, and $\ket{\SL\SL\SR}$ events. Ideally, this procedure prepared the state in Eq.~2.

To test Mermin's inequality, we rotated the QWP in front of the Pockels cell in each receiver from $0^{\circ}$ to $45^{\circ}$ so that the receivers measured either in the $R/L$ or $D/A$ bases. Over the course of the 1\unit{hr} 19\unit{min} experiment, Alice, Bob, and Charlie averaged single-photon detection rates of 78\unit{kHz}, 126\unit{kHz}, and 131\unit{kHz}, respectively with an average trigger rate of 505\unit{kHz}, giving a four-fold coincidence rate of 0.5\unit{Hz}. From these numbers we find that the free-space link efficiencies averaged 33\% and 32\% for Bob and Charlie, respectively, and the long fiber efficiency at Alice was 14\%.

\section*{Detailed Experimental Description}

\subsection*{Source of Three-Photon GHZ States}

\begin{figure*}[htbp]
    \centering
    \includegraphics[width=1.6\columnwidth]{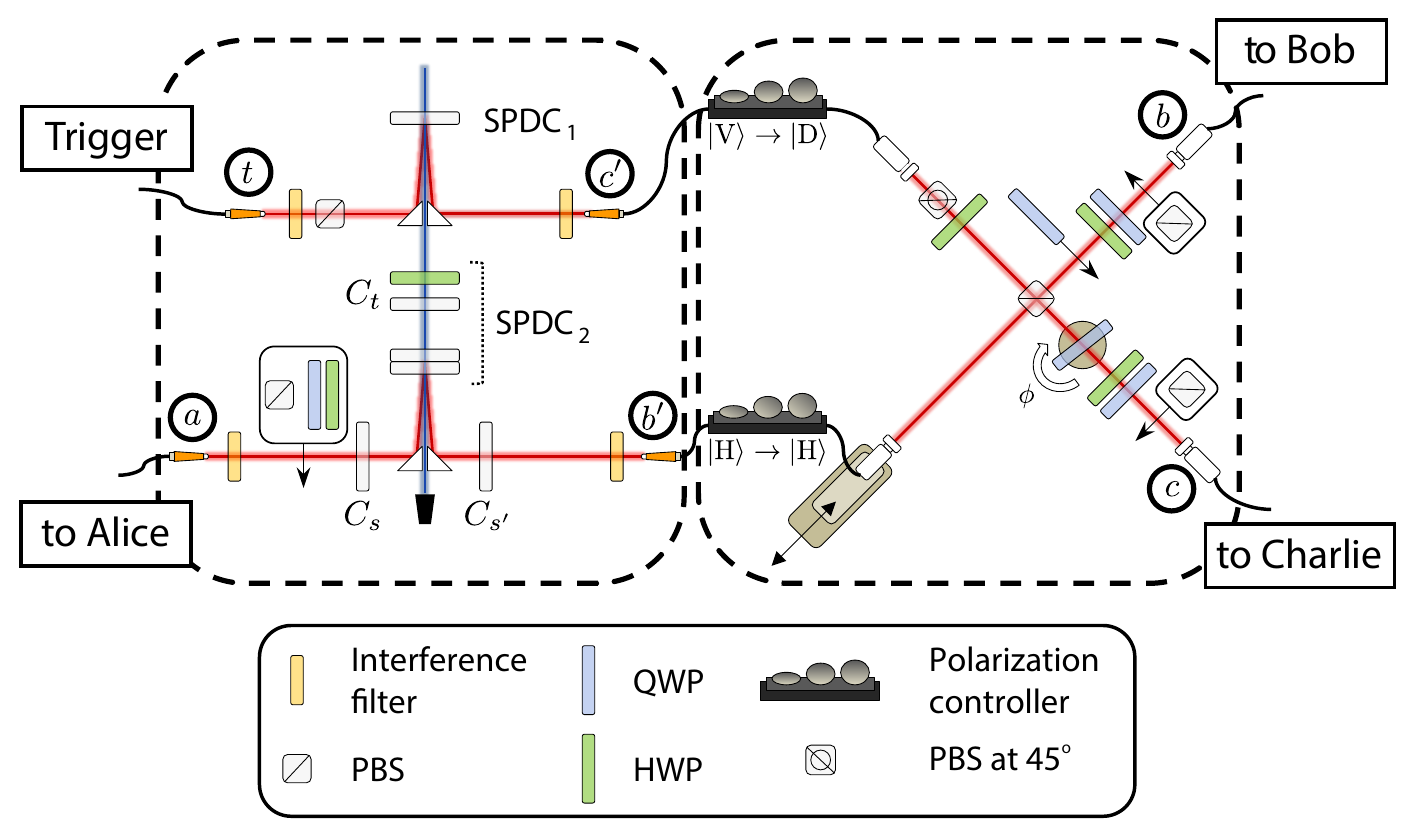}
    \caption{Two down-conversion sources ($\mathrm{SPDC_{1}}$, $\mathrm{SPDC_{2}}$) produced the state $\ket{\psi} \propto \ket{\SV_{t} \SV_{c'}} \otimes (\ket{\SH_{a} \SH_{b'}} + \ket{\SV_{a} \SV_{b'}})$ shared between the spatial modes \emph{t} (trigger), $a$, $b'$, and $c'$. Birefringent crystals ($C_{t}$, $C_{s}$, $C_{s'}$) were used to compensate temporal and transverse (spatial) walk-off in the second source. All photons went through interference filters and were coupled into single-mode fibers. Polarization controllers rotated the polarization of photons in mode $c'$ to diagonal, $\ket{\SD}$, while maintaining the polarization in mode $b'$ in its original state. The photons in modes $b'$ and $c'$ were combined on a polarizing beam-splitter. The delay in mode $b'$ was adjusted for optimum two-photon interference \cite{HOM87} while the QWP ($\phi$) controlled the phase. Photons in mode $b$ and $c$ were coupled into single-mode fibers and then sent to single photon detectors, a fourfold coincident event signalled a state of the form $\ket{\mathrm{GHZ}_{\pi}} = \frac{1}{\sqrt{2}} \left(\ket{\SH_{a} \SH_{b} \SH_{c}} +e^{i \pi} \ket{\SV_{a} \SV_{b} \SV_{c}} \right) \otimes \ket{\mathrm{trigger}}$. A QWP could be inserted and rotated in mode $b$ for further phase adjustments. Polarization analyzers consisting of a HWP, QWP, and a PBS could be inserted to characterize the quantum state locally.}
    \label{fig.Source}
\end{figure*}

Our source of triggered three-photon GHZ states was based on two consecutive spontaneous parametric down-conversion (SPDC) sources and an interferometer \cite{PDGWZ01}, as shown in Fig.~\ref{fig.Source}. The two sources were pumped with a pulsed titanium:sapphire laser (Spectra Physics Tsunami HP, repetition rate 80\unit{MHz}, average power 2.5\unit{W}, center wavelength 790\unit{nm}, and bandwidth 11.7\unit{nm}, FWHM) which was frequency doubled in a bismuth borate (BiBO) crystal to 395\unit{nm} (bandwidth 1.75\unit{nm}, and average power 810\unit{mW}). The first source ($\mathrm{SPDC_{1}}$) was pumped directly with the horizontally polarized ($\ket{\SH}$) up-converted beam focused onto a 1\unit{mm} thick $\beta$-barium borate ($\beta$-BBO) crystal cut for type-I non-collinear degenerate SPDC in order to produce vertically-polarized photon pairs at a wavelength of 790\unit{nm}. One of the photons, labeled \emph{t} for ``trigger'', passed through a polarizer and an interference filter (center wavelength 790\unit{nm}, bandwidth 3\unit{nm}), before being coupled into a single-mode fiber and detected with a single-photon detector. The other photon passed through a similar filter and was coupled into a single-mode fiber.

The pump pulse passed through a half-wave plate (HWP) followed by compensation crystals ($C_{t}$) before pumping the second SPDC source ($\mathrm{SPDC_{2}}$) made up of a pair of orthogonally oriented $\beta$-BBO crystals (also 1\unit{mm} thick type-I) used to create photon pairs in a Bell state \cite{LKR09}. The HWP rotated the pump polarization from horizontal to diagonal such that each pump photon was equally likely to produce a down-converted photon pair in either the first or the second BBO crystal. Compensation crystals eliminated temporal distinguishability between photons created in each BBO crystal and consisted of a 1\unit{mm} thick $\alpha$-BBO crystal and a 2\unit{mm} thick quartz crystal cut for maximum birefringence (together referred to as $C_{t}$ in Fig.~\ref{fig.Source}). Spatial walk-off was compensated in the first arm with a 1\unit{mm} thick BiBO ($C_{s}$) and in the second arm with two BiBO crystals ($C_{s'}$) totalling 1.25\unit{mm} thickness. These photons were also spectrally filtered and coupled into single-mode fibers. The combined state of the four photons at this point was of the form $\ket{\SV_{t} \SV_{c'}} \otimes \frac{1}{\sqrt{2}} (\ket{\SH_{a} \SH_{b'}} + e^{i \theta} \ket{\SV_{a} \SV_{b'}})$, with some phase, $\theta$, between the two terms of the Bell pair.

To produce the GHZ state, the photon in mode $b'$ was sent to a polarizing beamsplitter (PBS). The photon in mode $c'$ was rotated from vertical to diagonal polarization, passed through a diagonal polarizer, a HWP for fine-tuning and was sent to the PBS. The photons in modes $b'$ and $c'$ were made to temporally and spatially overlap at the PBS by adjusting the path length of mode $b'$ via a motorized translation stage. In mode $c$, a QWP with its fast axis vertically oriented was mounted on a motorized rotation stage to finely control the phase, $\phi$, of our state. Additionally, a QWP in mode $b$ could be inserted to introduce specific phase shifts (0, $\pi/2$, $\pi$, and $3\pi/2$). A four-fold coincidence detection in the four outputs $a$, $b$, $c$, and $t$ indicated the generation of the desired GHZ state
\begin{equation}\label{eq.GeneralGHZState}
  \ket{\mathrm{GHZ}_{\phi}} = \frac{1}{\sqrt{2}}(\ket{\SH_{a} \SH_{b} \SH_{c}} + e^{i\phi}\ket{\SV_{a} \SV_{b} \SV_{c}})
\end{equation}
in modes a, b, and c. With no analyzers in place we directly measured the singles rates of 576\unit{kHz}, 386\unit{kHz}, 417\unit{kHz}, and 476\unit{kHz} for modes \emph{a}, \emph{b}, \emph{c}, and \emph{t} respectively, and a four-fold coincidence rate of 39\unit{Hz}.

To completely characterize our state, we performed quantum state tomography immediately following the Mermin inequality experiments by inserting polarization analyzers in each arm, $a$, $b$, and $c$. Our polarization analyzers consisted of a HWP, followed by a QWP, and finally a PBS as shown in Fig.~\ref{fig.Source}. The photons transmitted through the PBS were coupled into single-mode fibers and sent to local single-photon counting modules and coincidence logic with a coincidence window of 6\unit{ns}.

\begin{figure}[htbp]
    \centering
    \includegraphics[width=0.75\columnwidth]{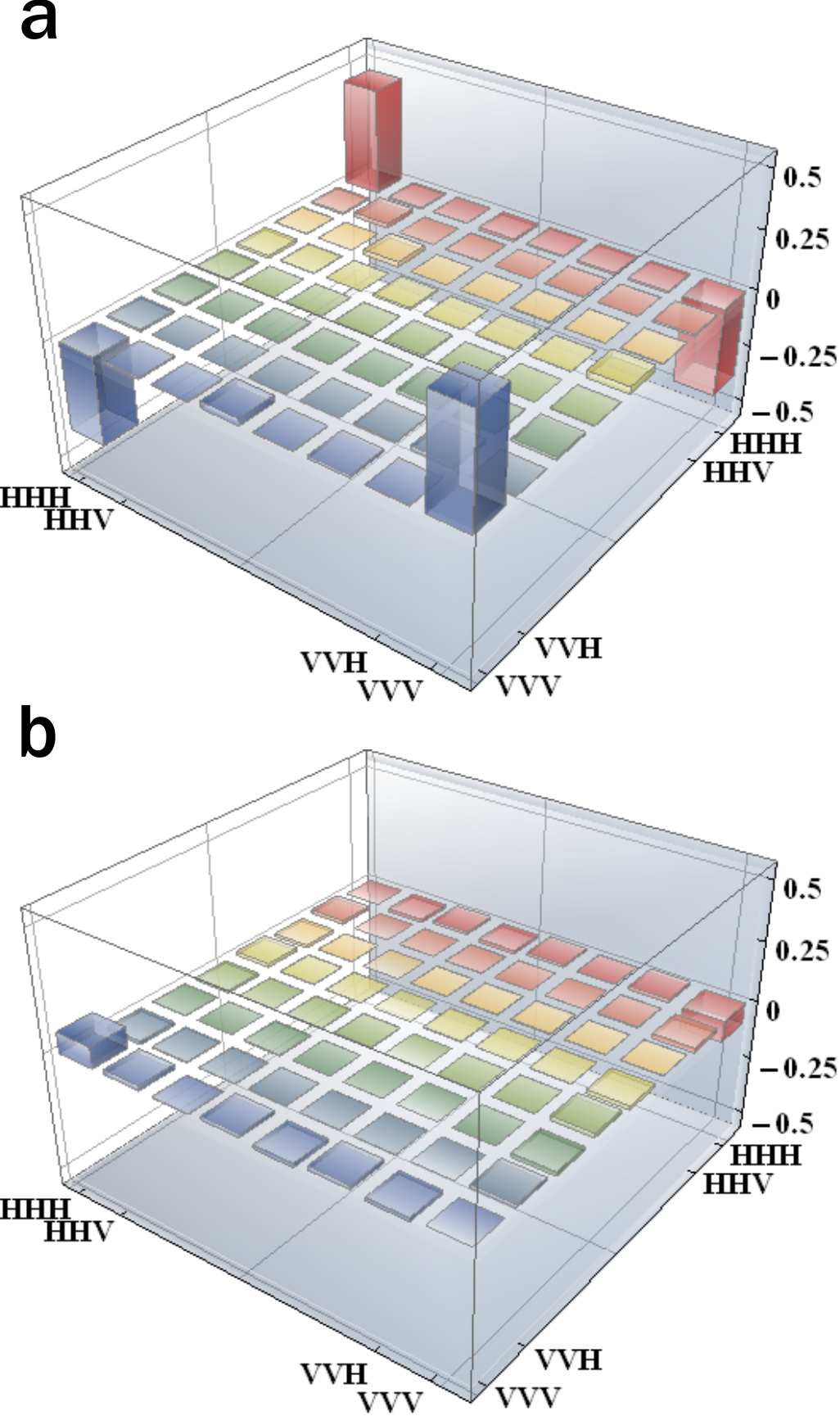}
    \caption{Quantum state tomography. Real (a) and imaginary (b) parts of the reconstructed density matrix of our GHZ state as measured directly in the laboratory. The fidelity of our state at the source with the desired state $\ket{\mathrm{GHZ}_{\pi}}$ was $(82.9 \pm 0.3)\%$.}
    \label{fig.Tomography}
\end{figure}

We used the measurement settings $\ket{\SH}$, $\ket{\SV}$, $\ket{\SD}$, $\ket{\SA}$, $\ket{\SR}$ and $\ket{\SL}$ for each photon, resulting in a total of 216 three-photon polarization measurements. We randomly cycled through all the measurement settings, counting for 5\unit{s} each, and repeated this procedure 10 times in order to average out fluctuations in the source. Using this data, we applied an iterative maximum-likelihood tomography algorithm \cite{JFH03} to reconstruct the state. The real and imaginary parts of the density matrix are shown in Fig.~\ref{fig.Tomography}. The fidelity with the desired state $\ket{\mathrm{GHZ_{\pi}}}$, where the phase was set to $\phi = \pi$, was $(82.9 \pm 0.3)\%$. The error on the fidelity is estimated using a Monte Carlo simulation. Note that the phase was adjusted to different settings using the QWPs during the Mermin inequality tests. From the reconstructed density matrix we calculate the highest possible Mermin parameter to be $M = 3.08$ for the optimal measurement settings.

%\begin{table}[htbp]
%  \centering
%  \begin{tabular}{|c|c|c|c|}
%    \hline
%    \textbf{Measurement} & \multirow{3}{*}{\textbf{Counts}} & \textbf{Measurement} & \multirow{3}{*}{\textbf{Counts}} \\
%    \textbf{settings} & & \textbf{settings} & \\
%    \textbf{(ABC)} & & \textbf{(ABC)} & \\
%    \hline
%    \textbf{HHH} & \textbf{775} & DDD & 60 \\
%    VHH & 19 & \textbf{ADD} & \textbf{339} \\
%    HVH & 37 & \textbf{DAD} & \textbf{370} \\
%    VVH & 23 & AAD & 45 \\
%    HHV & 31 & \textbf{DDA} & \textbf{384} \\
%    VHV & 34 & ADA & 42 \\
%    HVV & 12 & DAA & 66 \\
%    \textbf{VVV} & \textbf{765} & \textbf{AAA} & \textbf{340} \\
%    \hline
%    H/V Visibility & 81.60\% & D/A Visibility & 74.12\% \\
%    \hline
%  \end{tabular}
%  \caption{Subset of the counts from the quantum state tomography measurement.}
%  \label{tab.TomoCounts}
%\end{table}

\subsection*{Free-Space Receiver Configuration}

\begin{figure}[htbp]
    \centering
    \includegraphics[width=0.9\columnwidth]{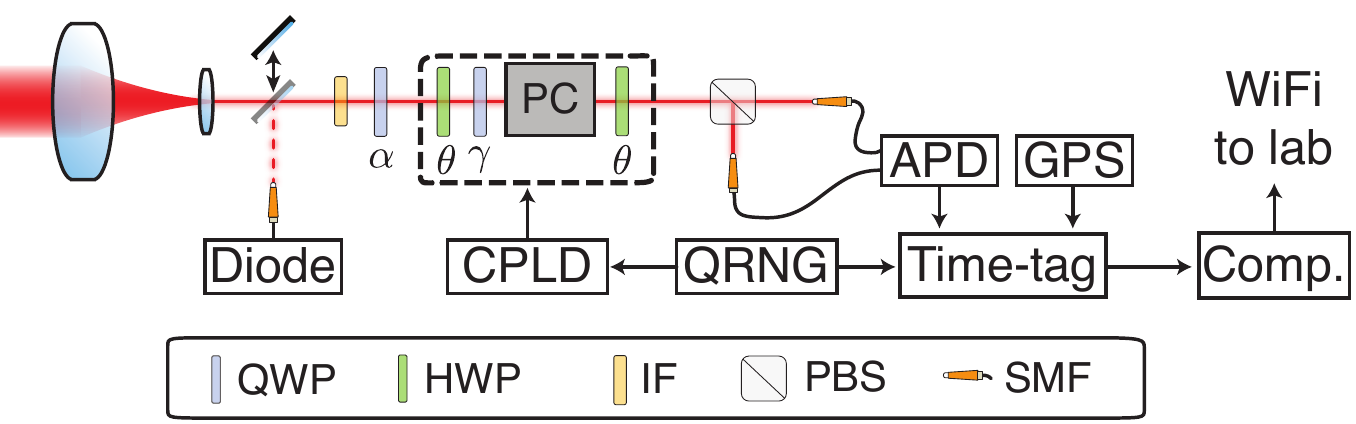}
    \caption{Schematic of the receiver telescope. Incoming light was focussed and collimated using a 150\unit{mm} telescope lens and 10\unit{mm} lens, before passing through a 10\unit{nm} interference filter (IF). In the dashed box, a Pockels' cell (PC), together with the QWP ($\gamma = 45^{\circ}$) and two HWPs ($\theta = -11.25^{\circ}$), acted as the identity when a ``0'' was received from the QRNG, or as a HWP at $22.5^{\circ}$ when a ``1'' was received. Before the dashed box, when the QWP was set at $\alpha = 0^{\circ}$ measurements were made in the \emph{H/V} and \emph{R/L} bases, for QRNG outputs and 0 and 1, respectively. When $\alpha = 45^{\circ}$, measurements were made in the \emph{R/L} and \emph{D/A} bases.}
    \label{fig.Receiver}
\end{figure}

A schematic of the free-space receivers employed at Bob and Charlie is shown in Fig.~\ref{fig.Receiver}. Alice used the same setup, without the initial telescope lenses and interference filter. A 150\unit{mm} diameter achromatic lens with a focal length of 400\unit{mm} (Optarius 12-7555) together with a second smaller lens (New Focus 5724-B-H 8.0 20x, diameter 9.9\unit{mm}, focal length 8.0\unit{mm}; or New Focus 5725-B-H 11.0 16x, diameter 7.2\unit{mm}, focal length 11\unit{mm}) collected and collimated the incoming beam for the polarization analyzer. A 790\unit{nm} bandpass interference filter with bandwidth 10\unit{nm} was used to suppress background. A flip-mounted mirror could be inserted along with a polarizer (not shown) to aid in alignment of the analyzer.

%\begin{figure*}[htbp]
%\end{figure*}

Each polarization analyzer contained a Pockels cell (Leysop Ltd, RTP-4-20-AR800-DMP) with a response time of 10\unit{ns}. The Pockels cell (PC) received a ``0'' or ``1'' signal from the QRNG and switched to the negative or positive of its quarter-wave voltage, respectively. Output from the QRNG was sampled at 1\unit{MHz} and was converted to a series of pulses using a complex programmable logic device (CPLD) electronic driver which switched the PC. The crystal in the PC had its optic axis at $45^{\circ}$, so that when it received a ``0''  (``1'') it acted as a QWP at $- 45^{\circ}$ ($+ 45^{\circ}$). We used a QWP with optic axis at $\gamma = 45^{\circ}$ directly before the PC so that it could be operated at its quarter-wave voltage rather than its half-wave voltage \cite{SUKRMHRFLJZ10}, enabling higher-speed switching. To calibrate the PC, we finely tuned its quarter-wave voltage, tip and tilt of the crystal, and angles of the QWP and HWPs. We measured contrasts of approximately 100:1 in each basis.

If the QRNG output was ``1'', then the QWP with $\gamma = 45^{\circ}$ and PC(``1'') acted as a HWP set to $45^{\circ}$. The HWPs before and after the QWP-PC combination were both set to $\theta = -11.25^{\circ}$ so that together the elements inside the dashed box acted as a HWP at $22.5^\circ$. A QWP ($\alpha$) was placed before this arrangement. By setting this QWP to $\alpha = 0^{\circ}$, which we did when setting the phase of the GHZ state, an incoming photon with polarization \emph{L} (\emph{R}) got mapped to \emph{H} (\emph{V}). This QWP was set to $\alpha = 45^{\circ}$ to test Mermin's inequality; for this setting, photons entering the receiver setup with polarization \emph{D} (\emph{A}) got mapped to \emph{H} (\emph{V}).

If the QRNG output a ``0'', then the QWP with $\gamma = 45^{\circ}$ and PC(``0'') acted as the identity operation. With the HWPs included, it still acted as the identity. When the QWP ($\alpha$) was set to $0^\circ$, incoming photons of polarization \emph{H} (\emph{V}) were unchanged. When $\alpha = 45^{\circ}$ the \emph{R} (\emph{L}) polarization was mapped to \emph{H} (\emph{V}). The projective measurement was made by a PBS, after which transmitted and reflected photons were collected into multi-mode fibers and detected with single photon detectors.

\subsection*{Quantum Random Number Generators}

We employed quantum random number generators based on the random splitting of photons on a beamsplitter \cite{JAWWZ00}. This particular method has the benefit that it provides random signals in a very brief time frame and does not introduce additional lag times from the post-processing of data. Each QRNG contained a green LED, a 50/50 beamsplitter, two photon counting photomultiplier tubes (PMT, Hamamatsu H10721P-110), and a flip-flop to hold the random bit. The device was packaged inside a nuclear instrumentation module.

The LED light passed through a pinhole and was split by the beamsplitter. Photons from each beam were detected using a PMT with counting threshold adjusted to output counts at a rate of 14\unit{MHz}. Both detectors were connected to the electronic flip-flop, such that the ``1'' detector set the flip-flop to state 1 and the ``0'' detector reset the flip-flop to state 0. The output of the flip-flop was then sampled by external logic at 1\unit{MHz}. Fine adjustment of the photomultiplier tube thresholds balanced the number of 0s and 1s to 500\unit{kHz} each with an accuracy of about $\pm 1\%$.

\begin{figure}[htbp]
    \centering
    \includegraphics[width=0.75\columnwidth]{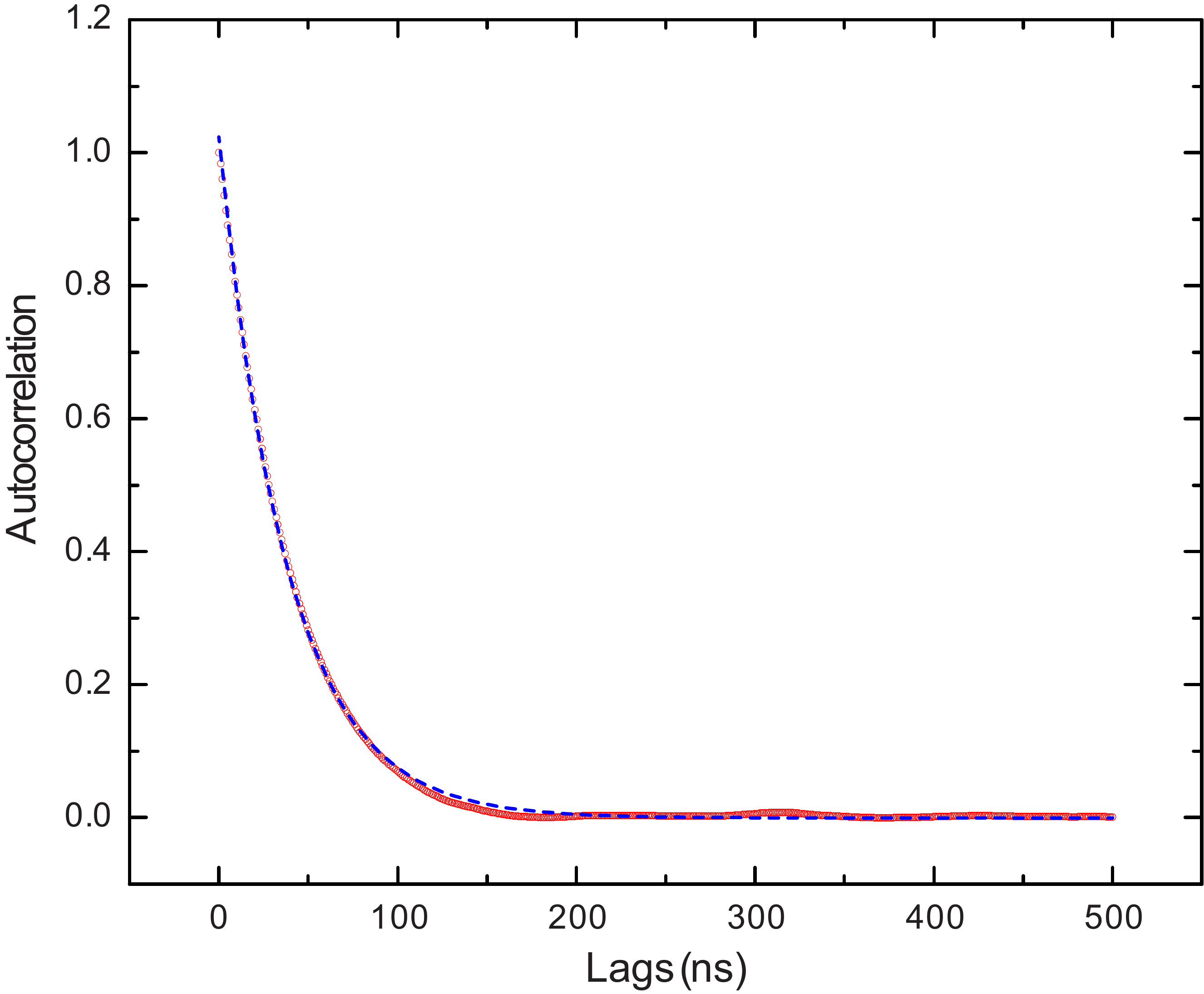}
    \caption{Typical measured autocorrelation data (red points) for our QRNGs where the measured autocorrelation time was $\tau = 38.35\unit{ns}$ as determined by an exponential fit (blue dotted line). For our location and timing model, we consider an autocorrelation window of 200\unit{ns}, ensuring uncorrelated values at all prior times.}
    \label{fig.QRNGAutoCorrelation}
\end{figure}

We directly measured the output of each QRNG before the 1\unit{MHz} sampling. From these data, we calculated the autocorrelation function of each QRNG. Fig.~\ref{fig.QRNGAutoCorrelation} shows a typical autocorrelation function for one of our QRNGs with an exponential fit to the data giving a $1/e$ autocorrelation time of 38\unit{ns}. This characterizes the timescale over which the QRNG output is predictable. To account for this predictability over short times, we included a \emph{QRNG autocorrelation} time factor in our space-time analysis of 200\unit{ns}, over 5 times the $1/e$ timescale. Furthermore, the sampling rate of the QRNG at 1\unit{MHz} or every 1$\unit{\mu s}$ was on a much longer timescale than the autocorrelation time, thus subsequent output bits are uncorrelated.

Internal hardware delays of $33 \pm 3\unit{ns}$, $34 \pm 3\unit{ns}$, and $37 \pm 3\unit{ns}$  were measured for the QRNGs at Alice, Bob, and Charlie respectively by pulsing the light-emitting diode and recording the time when the flip-flop switched. These measured delays are incorporated into the space-time analysis.

During the experiment we monitored the performance of the QRNGs in two ways. First, we measured the relative bias between ones and zeros in the QRNG output. Second, we recorded sequences of repeated bits, blocks of 0's or 1's with exactly $n$ length, where we include block lengths from $n = 1$ to $n = 16$, for each second of QRNG output. We calculated the $\chi^2$ parameter \cite{Knu69} to quantify how far the output of each QRNG was from the distribution of block sizes expected from a truly random number generator, which is expressed as
\begin{equation}
\chi^2 = \sum_{i=1}^{16} \frac{ (P^i_{0} - P^i_{0,\mathrm{th}})^2  }{ P^i_{0,\mathrm{th}} }+  \frac{ (P^i_{1} - P^i_{1,\mathrm{th}})^2}{ P^i_{1,\mathrm{th}}}.
\end{equation}
Here $P^i_{k}$ is the measured probability of finding a string of length $i$ comprised entirely of bit value $k$ in the given QRNG output; $P^i_{k,\mathrm{th}}$ is the probability of finding a string of length $i$ of entirely bit $k$ in a truly random output. The QRNGs were adjusted to optimize their performance. Over the course of the experiment Alice, Bob, and Charlie measured average $\chi^{2}$ values of 33.0, 41.1, and 68.1 respectively, assuming equal probabilities for 0 and 1. We also calculated $\chi^{2}$ values for each location taking into account the bias of the QRNGs. The average values of this biased $\chi^{2}$ parameter for Alice, Bob, and Charlie were 30.5, 34.8, and 48.0 respectively.

The basis selection made at Randy was transmitted to Alice via a free-space RF link. Electronic pulses output from the QRNG and CPLD logic at Randy were converted from digital to analog signals using a commercial video electronics transceiver, and sent to the RAC building using a 2.4\unit{GHz} parabolic antenna. An identical receiving antenna was placed outside the RAC building, along with a second transceiver to digitize the signal. A CPLD logic unit at Alice, connected to the RF receiver via a coaxial cable, sampled the incoming signal at a rate of 80\unit{MHz}, and output pulses for the Pockels cell whenever it detected a change in basis. Finally, since Alice's random numbers were generated at Randy, we recorded the random numbers at both locations in order to verify the faithful transmission. Over the course of the experiment 4,450,916,576 random bits generated at Randy were correctly sent to Alice without a single error.

\subsection*{Electronics and Software}

All single-photon detection events were recorded using time-tagging electronics (156.25\unit{ps} precision) with GPS timing units (Spectrum Instruments TM-4) to provide a 10\unit{MHz} reference signal. A one pulse-per-second (1PPS) signal from the GPS units was also recorded along with detailed date and time information via an RS-232 interface in order to aid synchronization and coincidence searching. We simultaneously recorded time-tags for each QRNG both to monitor their performance and correlate detection events with measurement settings. With each time-tag saved as a double-precision integer (64 bits) and measurement result saved as a byte (8 bits), this produced data files that were approximately 48\unit{GB} in size at each of the four locations over the course of the experiment detailed in the main text.

To set the phase of the GHZ state and monitor the experiment during data acquisition, measured time-tags were transmitted over a long-distance wireless network to the computer at RAC. The wireless network supported upload rates of 30\unit{Mbps} for each of Bob and Charlie and also allowed remote operation of their computers from RAC. Custom software correlated detection events and measurement settings using time-tags, then found the two-fold coincidence counts between Alice \& Bob and Alice \& Charlie and the appropriate offset delays, and then finally the four-fold coincidences including the trigger detection events.

%In order to achieve real time analysis of the Mermin parameter, classical time tag information needed to be sent from Bob and Charlie back to Alice as it was recorded. To obtain this, a Ubiquiti AirMax M series wireless system was implemented. At each Bob and Charlie were a NanoStation M2 sender/receiver and at the dome for receiving was a Rocket M2 with an 120 degree AirMax Sector antenna. With this system we were able to have 30Mbps upload rate from each Bob and Charlie and allowed for real time analysis as well as observation of the phase drift.

\section*{Space-Time Analysis}

\subsection*{Locations and Delays}

The locations in the experiment were measured using a GPS timing unit, integrating signals over about 10\unit{min} at each location. The coordinates of Bob, Charlie, Randy, the RF receiver outside the RAC building, and the rooftop dome on RAC were directly measured. We could not obtain a GPS signal inside the RAC building, so instead measured the four corners of the building and calculated the position of Alice and the Source based on building schematics. Elevations were determined using city and regional data \cite{ElData}. The coordinates are shown in Table~\ref{tab.GPSCoords}. We convert these into 3-dimensional Cartesian coordinates and use an uncertainty of 5\unit{m} in each dimension for our calculations, based on the manufacturer's specifications of the GPS unit.

\begin{table}[htbp]
 \centering
 \begin{tabular}{|l|r|r|r|}
  \hline
  Location & Lat.\ ($^\circ$ N) & Long.\ ($^\circ$ W) & El.\ (m) \\
  \hline
  Bob & 43.477495 & 80.564536 & 347.0 \\
  Charlie & 43.483830 & 80.560289 & 355.0 \\
  Randy & 43.478300 & 80.549260 & 349.0 \\
  Dome & 43.478907 & 80.555169 & 356.8 \\
  RAC RF receiver & 43.478717 & 80.554490 & 344.5 \\
  RAC SE corner & 43.478623 & 80.554445 & 341.5 \\
  RAC SW corner & 43.478516 & 80.554976 & 341.5 \\
  RAC NW corner & 43.478941 & 80.555278 & 341.5 \\
  RAC NE corner & 43.479114 & 80.554750 & 341.5 \\
  Alice/Source (calc.) & 43.478737 & 80.554759 & 342.5 \\
  \hline
 \end{tabular}
 \caption{GPS coordinates of the various locations relevant to the experiment.}
 \label{tab.GPSCoords}
\end{table}

Our space-time analysis incorporates all delays experienced by the photons, starting from their creation in the source, until they are ultimately detected and recorded in the time-tagging electronics of Alice, Bob, or Charlie. Table~\ref{tab.PhotonDelays} shows the delays for a photon that arrives at each site. For each path, we measured the path length in the source, labelled \emph{Source Path}. In Alice's case this was the free-space distance from the down-conversion crystal to the fiber coupler; for Bob and Charlie's cases, this included free-space distance from the down-conversion crystal to the first fiber couplers, the short fibers into and out of the HOM interferometer, and the free-space distance inside the interferometer to just before the long fibers to the roof top. We separately measured the optical delay through the long fiber spool for Alice, labelled \emph{Delay fiber}, plus a short free-space path before the Pockels cell. We measured the delay through the long fibers from the lab to the roof top for Bob and Charlie, labelled \emph{Fiber to roof and telescope}. This delay includes the telescope itself. Next we calculated the free-space delay based on the measured free-space distance from the telescope to the Pockels' cell in each receiver, labelled \emph{Free-space}. Finally, we measured the optical delay from the Pockels cell to the detector, the detector and electronic delay, estimated (based on cable length and internal clock frequency) the time needed to enter the time-tagger memory; this is labelled \emph{Measurement}.

\begin{table}[htbp]
  \centering
  \begin{tabular}{|c|c|}
      \hline
      \multicolumn{2}{|c|}{Alice's Photon Delays (ns)} \\
      \hline
      Source path & $3.39 \pm 0.04$ \\
      Delay fiber & $2875.7 \pm 0.7$ \\
      Measurement & $47.2 \pm 5.1$ \\
      \textbf{Total} & $2926.3 \pm 5.1$ \\
      \hline
      \multicolumn{2}{|c|}{Bob's Photon Delays (ns)} \\
      \hline
      Source path & $34.2 \pm 0.9$ \\
      Fiber to roof and telescope & $420.2 \pm 3.0$ \\
      Free-space & $2577.1 \pm 24$ \\
      Measurement & $47.2 \pm 5.1$ \\
      \textbf{Total} & $3078.7 \pm 24$ \\
      \hline
      \multicolumn{2}{|c|}{Charlie's Photon Delays (ns)} \\
      \hline
      Source path & $34.2 \pm 0.9$\\
      Fiber to roof and telescope & $420.5 \pm 3.0$ \\
      Free-space & $2289.6 \pm 24$ \\
      Measurement & $47.2 \pm 5.1$ \\
      \textbf{Total} & $2791.5 \pm 24$ \\
      \hline
  \end{tabular}
  \caption{Time delays for the space-time analysis.}
  \label{tab.PhotonDelays}
\end{table}

Our model also requires the times at which the measurement basis choices are made. Because the source, QRNGs, and receiver logic were not synchronized, this does not occur at a specific time relative to the source creation, but rather occurs over some range of times. We thus require the earliest and latest times at which each basis choice was made at each location.

A summary of the important times is shown in Table~\ref{tab.BasisSelectionDelays}. We begin by describing the time of the measurement settings at Bob and Charlie. The time from an LED photon in the QRNG hitting the beamsplitter to the generation of electronic output pulses for the Pockels cell is labelled \emph{QRNG logic}. These pulses trigger the Pockels cell after the delay labelled \emph{PC}. The sampling of the QRNGs were not synchronized to each other, or to the source. To account for this, we incorporate an additional basis sampling period of $1\unit{\mu s}$ to the maximum time before the measurement that the QRNG could have set the basis; this is labelled \emph{QRNG asynchronicity}. We also add an additional autocorrelation factor to this maximum time labelled \emph{QRNG autocorrelation} as described earlier to account for QRNG predictability over short times. The CPLD electronic driver contained an electronic delay written into its firmware, labelled \emph{Extra programmed delay}. Together with the chosen locations, this programmable delay was used to optimize the loophole closure tolerances. Adding these delays up, we find that Bob's basis selection occurred at minimum $665 \pm 6$\unit{ns} and at maximum $1865 \pm 6$\unit{ns} before his Pockels cell was triggered. Similarly, Charlie's basis selection occurred at minimum $663 \pm 6$\unit{ns} and at maximum $1863 \pm 6$\unit{ns} before his cell was triggered. To place these events in the space-time diagrams in Fig.3, we start from the total photon delay from Table~\ref{tab.PhotonDelays} and subtract the measurement time; this is the delay from the source to a Pockels cell. From this we subtract the maximum total basis selection delay from Table~\ref{tab.BasisSelectionDelays} to find the earliest time at which the basis was selected, labelled for example \emph{Bob basis early} in Fig.~3. We can subtract instead the minimum total basis delay to find the latest time, labelled for example \emph{Bob basis late}.

\begin{table}[htbp]
  \centering
  \begin{tabular}{|c|c|}
      \hline
      \multicolumn{2}{|c|}{Alice's Basis Selection Delays (ns)} \\
      \hline
       \multirow{2}{*}{QRNG autocorrelation} & 0 (min.) \\
      & 200 (max.) \\
      \multirow{2}{*}{QRNG asynchronicity} & 0 (min.) \\
       & 1000 (max.) \\
      QRNG logic & $72 \pm 4$ \\
      RF free-space & $1840 \pm 24$ \\
      Coaxial cable & $140 \pm 3$ \\
      \multirow{2}{*}{RF receiver logic async.\ and delay} & $26 \pm 2$ (min.) \\
       & $38.5 \pm 2$ (max.) \\
      PC & $140 \pm 5$ \\
      \multirow{2}{*}{\textbf{Total}} & $2218 \pm 25$ (min.) \\
       & $3431 \pm 25$ (max.) \\
      \hline
      \multicolumn{2}{|c|}{Bob's Basis Selection Delays (ns)} \\
      \hline
      \multirow{2}{*}{QRNG autocorrelation} & 0 (min.) \\
      & 200 (max.) \\
      \multirow{2}{*}{QRNG asynchronicity} & 0 (min.) \\
       & 1000 (max.) \\
      QRNG logic & $50 \pm 4$ \\
      Extra programmed delay & 475 \\
      PC & $140 \pm 5$ \\
      \multirow{2}{*}{\textbf{Total}} & $665 \pm 6$ (min.) \\
       & $1865 \pm 6$ (max.) \\
      \hline
      \multicolumn{2}{|c|}{Charlie's Basis Selection Delays (ns)} \\
      \hline
      \multirow{2}{*}{QRNG autocorrelation} & 0 (min.) \\
      & 200 (max.) \\
      \multirow{2}{*}{QRNG asynchronicity} & 0 (min.) \\
       & 1000 (max.) \\
      QRNG logic & $48 \pm 4$ \\
      Extra programmed delay & 475 \\
      PC & $140 \pm 5$ \\
      \multirow{2}{*}{\textbf{Total}} & $663 \pm 6$ (min.) \\
       & $1863 \pm 6$ (max.) \\
      \hline
  \end{tabular}
  \caption{Measurement basis choice delays. Minimum and maximum total delays correspond to the earliest and latest times that events may occur, respectively.}
  \label{tab.BasisSelectionDelays}
\end{table}

The time of the measurement settings for Alice has some additional contributions as it is made at another location: Randy. Here, the \emph{QRNG logic} delay measures the time between an LED photon hitting the beamsplitter to the generation of an electronic output pulse which will then be transmitted over the RF link. This is longer than for Bob and Charlie due to the delay from one extra clock cycle in Randy's CPLD electronic driver. The time for an electronic signal to pass over the free-space RF link, including the transmitting and receiving electronics, is labelled \emph{RF free-space}. On the receiving end of the link, the signal passes from outside RAC to Alice by a \emph{Coaxial cable}. The signal passes to the RF receiver logic, a CPLD electronic driver, which samples the signal at 80\unit{MHz} and outputs electronic pulses for Alice's Pockel's cell. We account for additional asynchronicity due to this receiver logic by adding another 12.5$\unit{ns}$ to the maximum time for the basis selection; this is labelled \emph{RF receiver logic async.\ and delay}.

\subsection*{Tolerances for Loophole Closures }

We first calculate the minimum temporal separation between the relevant events which show our closure of the freedom-of-choice loophole using the values in Tables~\ref{tab.PhotonDelays} and \ref{tab.BasisSelectionDelays}, and the distances between experiment locations from Table~\ref{tab.GPSCoords}. Working back from when Bob detects a photon, we calculate his latest possible basis choice correlated to this detection as:
\begin{eqnarray}
3078.7\unit{ns} & \quad & \textrm{(Bob detection)} \nonumber \\
- 47.2\unit{ns}  &\quad & \textrm{(Measurement)} \nonumber \\
- 140\unit{ns}  & \quad & \textrm{(Pockel's cell)} \nonumber \\
- 475\unit{ns} & \quad & \textrm{(Extra prog. delay)} \nonumber \\
- 50\unit{ns} & \quad & \textrm{(QRNG logic)} \nonumber \\
= 2366.5\unit{ns}.
\end{eqnarray}
The time a signal at the speed of light takes to travel from the photon source to Bob is
\begin{equation}
\frac{800.6\unit{m}}{0.299792\unit{m/ns}} = 2670.5\unit{ns}.
\end{equation}
Therefore, the tolerance for the freedom-of-choice loophole closure is $2670.5\unit{ns} - 2366.5\unit{ns} = 304\unit{ns}$, shown as a red arrow in Fig.~3~(b). Thus, a signal carrying information about the source state generation, travelling at the speed of light, is $304\unit{ns}$ too late to affect Bob's choice of measurement basis.

Next we calculate the tolerance for the locality loophole closure. This occurs in the space-time slice containing Randy and Charlie in Fig.~3~(d). The earliest possible basis choice from Randy's QRNG can be worked back from Alice's detection event:
\begin{eqnarray}
2926.3\unit{ns} & \quad & \textrm{(Alice detection)} \nonumber \\
- 47.2\unit{ns} &\quad & \textrm{(Measurement)} \nonumber \\
- 3431\unit{ns} & \quad & \textrm{(Alice's maximum basis selection delay)} \nonumber \\
= -551.9\unit{ns}.
\end{eqnarray}
The earliest basis choice at Randy is $551.9\unit{ns}$ before the source produces the photons. The time a signal travelling at the speed of light takes to arrive at Charlie from Randy is
\begin{equation}
  -551.9\unit{ns} + \frac{1081.45\unit{m}}{0.299792\unit{m/ns}} = 3055.4\unit{ns}
\end{equation}
where the distance between Randy and Charlie ($1081.45\unit{m}$) can be calculated from Fig.~3~(a). Therefore, the tolerance for the locality loophole closure is $3055.4\unit{ns} - 2791.5\unit{ns} = 263.9\unit{ns}$, as shown in red in Fig.~3~(d).

\section*{Additional Experimental Results}

\subsection*{Main Text Mermin Inequality Results}

To support the data shown in the main text, we provide the raw counts measured in the 1\unit{hr} 19\unit{min} experiment. These are shown in Table~\ref{tab.MerminCounts} sorted by the 64 different possible measurement combinations. From these results we can extract 8 correlations, four of which were used to calculate Mermin's inequality in the main text. The remaining four can also be used to test Mermin's inequality keeping the measurement settings the same, such that $\mathbf{a}$, $\mathbf{b}$, and $\mathbf{c}$ correspond to $R/L$ and $\mathbf{a'}$, $\mathbf{b'}$, and $\mathbf{c'}$ to $D/A$, but writing the inequality in the (equally valid) form,
\begin{eqnarray}
    M' &=& | E(\mathbf{a'},\mathbf{b'},\mathbf{c'}) -  E(\mathbf{a},\mathbf{b},\mathbf{c'}) -  \nonumber \\
     && E(\mathbf{a},\mathbf{b'},\mathbf{c}) - E(\mathbf{a'},\mathbf{b},\mathbf{c}) | \leq 2. \label{eq.MerminInequality2}
\end{eqnarray}

\begin{table}[htbp]
  \centering
  \begin{tabular}{|c|c|c|c|c|c|c|c|}
      \hline
      \multicolumn{8}{|c|}{Counts (Alice,Bob,Charlie)} \\
      \hline
      \textbf{RRR} & \textbf{RRL} & \textbf{RLR} & \textbf{RLL} & \textbf{LRR} & \textbf{LRL} & \textbf{LLR} & \textbf{LLL} \\
      \hline
      55 & 9 & 18 & 92 & 9 & 53 & 77 & 15 \\
      \hline
      \textbf{RRD} & \textbf{RRA} & \textbf{RLD} & \textbf{RLA} & \textbf{LRD} & \textbf{LRA} & \textbf{LLD} & \textbf{LLA} \\
      \hline
      14 & 49 & 55 & 40 & 32 & 30 & 35 & 60 \\
      \hline
      \textbf{RDR} & \textbf{RDL} & \textbf{RAR} & \textbf{RAL} & \textbf{LDR} & \textbf{LDL} & \textbf{LAR} & \textbf{LAL} \\
      \hline
      21 & 46 & 72 & 43 & 34 & 30 & 23 & 50 \\
      \hline
      \textbf{RDD} & \textbf{RDA} & \textbf{RAD} & \textbf{RAA} & \textbf{LDD} & \textbf{LDA} & \textbf{LAD} & \textbf{LAA} \\
      \hline
      3 & 53 & 68 & 15 & 43 & 9 & 13 & 72 \\
      \hline
      \textbf{DRR} & \textbf{DRL} & \textbf{DLR} & \textbf{DLL} & \textbf{ARR} & \textbf{ARL} & \textbf{ALR} & \textbf{ALL} \\
      \hline
      21 & 39 & 56 & 55 & 37 & 26 & 28 & 59 \\
      \hline
      \textbf{DRD} & \textbf{DRA} & \textbf{DLD} & \textbf{DLA} & \textbf{ARD} & \textbf{ARA} & \textbf{ALD} & \textbf{ALA} \\
      \hline
      3 & 61 & 80 & 15 & 51 & 16 & 14 & 100  \\
      \hline
      \textbf{DDR} & \textbf{DDL} & \textbf{DAR} & \textbf{DAL} & \textbf{ADR} & \textbf{ADL} & \textbf{AAR} & \textbf{AAL} \\
      \hline
      8 & 56 & 69 & 18 & 34 & 8 & 14 & 81 \\
      \hline
      \textbf{DDD} & \textbf{DDA} & \textbf{DAD} & \textbf{DAA} & \textbf{ADD} & \textbf{ADA} & \textbf{AAD} & \textbf{AAA} \\
      \hline
      30 & 28 & 32 & 59 & 16 & 34 & 30 & 56 \\
      \hline
      \multicolumn{4}{|c|}{$M$} & \multicolumn{4}{|c|}{$M'$} \\
      \hline
      \multicolumn{4}{|c|}{$2.7720 \pm 0.0824$} & \multicolumn{4}{|c|}{$0.7746 \pm 0.1108$} \\
      \hline
  \end{tabular}
  \caption{The counts for all 64 different measurement possibilities used to measure the Mermin parameter.}
  \label{tab.MerminCounts}
\end{table}

\begin{figure}[htbp]
    \centering
    \includegraphics[width=0.95\columnwidth]{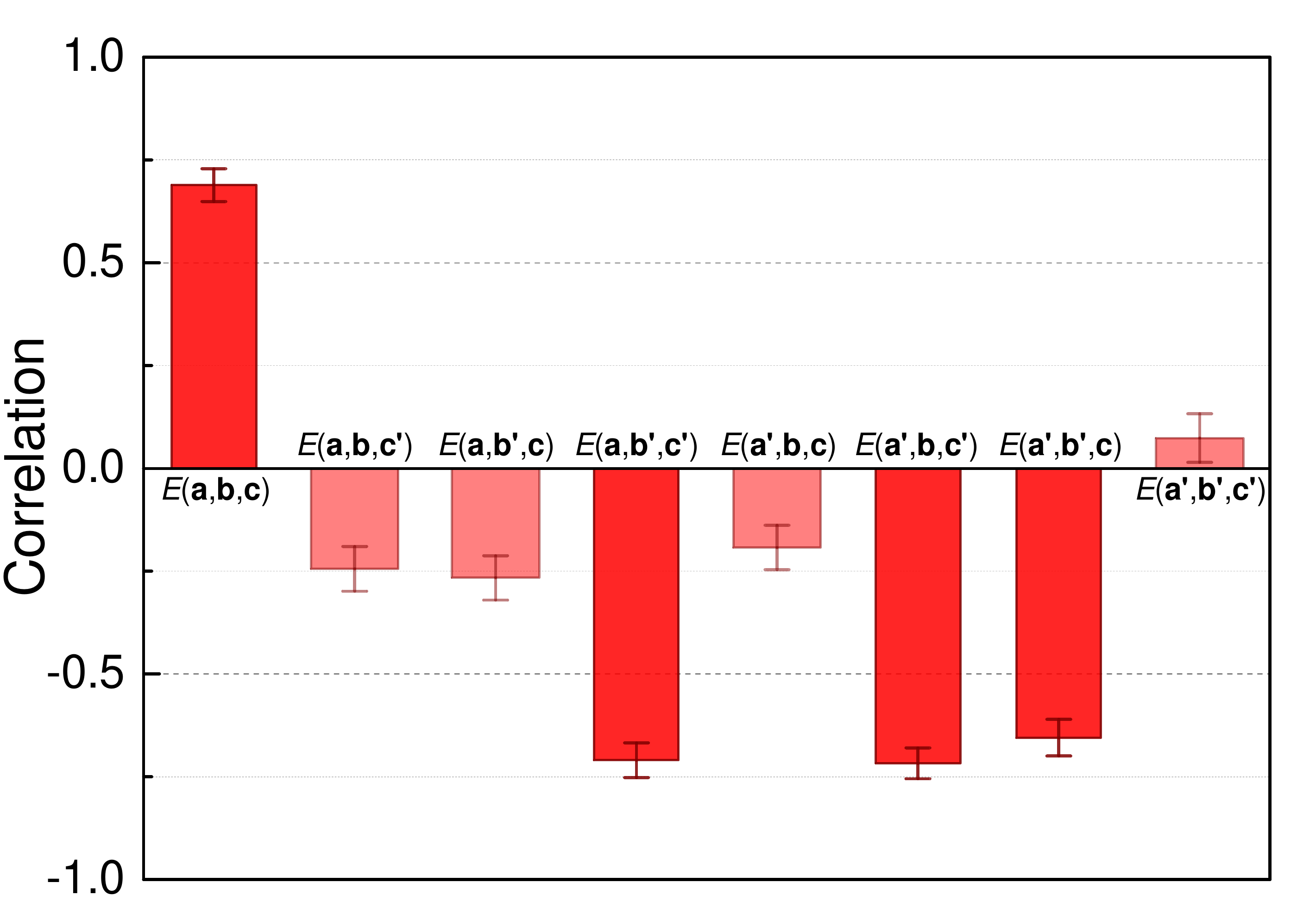}
    \caption{The complete set of eight experimentally measured three-photon correlations $E(\mathbf{a},\mathbf{b},\mathbf{c}) = 0.6890 \pm 0.0400$, $E(\mathbf{a},\mathbf{b},\mathbf{c'}) = -0.2444 \pm 0.0546$, $E(\mathbf{a},\mathbf{b'},\mathbf{c}) = -0.2665 \pm 0.0540$, $E(\mathbf{a},\mathbf{b'},\mathbf{c'}) = -0.7101 \pm 0.0424$, $E(\mathbf{a'},\mathbf{b},\mathbf{c}) = -0.1900 \pm 0.0538$, $E(\mathbf{a'},\mathbf{b},\mathbf{c'}) = -0.7176 \pm 0.0378$, $E(\mathbf{a'},\mathbf{b'},\mathbf{c}) = -0.6552 \pm 0.0444$, and $E(\mathbf{a'},\mathbf{b'},\mathbf{c'}) = 0.0737 \pm 0.0591$ (see Table~\ref{tab.MerminCounts} for raw count data). Error bars represent 1 standard deviation based on Poisson statistics.}
    \label{fig.ExpectationsFull}
\end{figure}

We show all 8 correlations in Fig.~\ref{fig.ExpectationsFull}. The four correlations used in the main text are shown in red and strongly violate Mermin's inequality with a Mermin parameter of $M = 2.77 \pm 0.08$; the remaining four yield a Mermin parameter $M' = 0.77 \pm 0.11$ and do not show a violation, as expected.

\subsection*{Additional Phases}

In addition to the Mermin inequality violation discussed in the main text, we performed additional tests where the three phases ($\phi \in \{+\pi/2, -\pi/2, 0\}$ in Eq.~\ref{eq.GeneralGHZState}) for the GHZ state were prepared. In each case, we measured data for approximately 30\unit{min}. The relevant correlations corresponding to a Mermin inequality for each state are summarized in Table~\ref{tab.AdditionalPhasesExpectationsAndMermin}. For all cases we measured a significant violation of Mermin's inequality. Note that the measured correlation values change sign when we change the GHZ state phase from $\phi = -\pi/2$ to $\phi = \pi/2$ as expected. When the phase is set to $\phi = 0$, we use the four correlations comprising $M'$ to show a violation.

\begin{table}[htbp]
  \centering
  \begin{tabular}{|c|c|c|c|}
      \hline
      \multicolumn{4}{|c|}{\textbf{Experiment I ($\ket{\textrm{GHZ}_{-\frac{\pi}{2}}}$)}} \\
      \hline
      $E(\mathbf{a},\mathbf{b},\mathbf{c})$ & $E(\mathbf{a},\mathbf{b'},\mathbf{c'})$ & $E(\mathbf{a'},\mathbf{b},\mathbf{c'})$ & $E(\mathbf{a'},\mathbf{b'},\mathbf{c})$ \\
      \hline
      $0.766 \pm 0.053$ & $-0.692 \pm 0.060$ & $-0.656 \pm 0.068$ & $-0.657 \pm 0.065$ \\
      \hline
      \multicolumn{4}{|c|}{$M$} \\
      \hline
      \multicolumn{4}{|c|}{$2.770 \pm 0.123$} \\
      \hline
      \multicolumn{4}{|c|}{\textbf{Experiment II ($\ket{\textrm{GHZ}_{\frac{\pi}{2}}}$)}} \\
      \hline
      $E(\mathbf{a},\mathbf{b},\mathbf{c})$ & $E(\mathbf{a},\mathbf{b'},\mathbf{c'})$ & $E(\mathbf{a'},\mathbf{b},\mathbf{c'})$ & $E(\mathbf{a'},\mathbf{b'},\mathbf{c})$ \\
      \hline
      $-0.631\pm 0.068$ & $0.664 \pm 0.063$ & $0.577 \pm 0.074$ & $0.664 \pm 0.061$ \\
      \hline
      \multicolumn{4}{|c|}{$M$} \\
      \hline
      \multicolumn{4}{|c|}{$2.537 \pm 0.134$} \\
      \hline
      \multicolumn{4}{|c|}{\textbf{Experiment III ($\ket{\textrm{GHZ}_{0}}$)}} \\
      \hline
      $E(\mathbf{a},\mathbf{b},\mathbf{c'})$ & $E(\mathbf{a},\mathbf{b'},\mathbf{c})$ & $E(\mathbf{a'},\mathbf{b},\mathbf{c})$ & $E(\mathbf{a'},\mathbf{b'},\mathbf{c'})$ \\
      \hline
      $-0.603 \pm 0.074$ & $-0.649 \pm 0.063$ & $-0.567 \pm 0.075$ & $0.672 \pm 0.064$ \\
      \hline
      \multicolumn{4}{|c|}{$M'$} \\
      \hline
      \multicolumn{4}{|c|}{$2.490 \pm 0.138$} \\
      \hline
  \end{tabular}
  \caption{The relevant measured correlations and Mermin parameter for the states $\ket{\mathrm{GHZ}_{-\frac{\pi}{2}}}$, $\ket{\mathrm{GHZ}_{\frac{\pi}{2}}}$, and $\ket{\textrm{GHZ}_{0}}$.}
  \label{tab.AdditionalPhasesExpectationsAndMermin}
\end{table}

\subsection*{Random to Deterministic Basis Choices}

We examined the effect of switching from random basis choices made with the QRNGs to deterministic, periodic choices made with function generators on the measured Mermin parameter. In each case we integrated measurements for approximately 15\unit{min}. We began with the source set to produce $\ket{\mathrm{GHZ}_{-\frac{\pi}{2}}}$ states and the QRNGs used to choose the measurement bases. The appropriate measured correlations used to calculate Mermin's inequality are shown in the top half of Table~\ref{tab.DeterministicBasisChoicesExpectationsAndMermin}.

Then Alice, Bob, and Charlie used deterministic function generators, set to output a periodic 500\unit{kHz} square wave, to choose their measurement bases. During the experiment, there was a problem with Charlie's PC and data had to be combined from two shorter successful measurements, resetting the PC electronics in between. The measured correlations for this case are shown in the bottom half of Table~\ref{tab.DeterministicBasisChoicesExpectationsAndMermin}. From these measurements we see the Mermin's inequality is violated regardless of whether the switching is random or periodic.

\begin{table}[htbp]
  \centering
  \begin{tabular}{|c|c|c|c|}
      \hline
      \multicolumn{4}{|c|}{\textbf{Experiment I (random)}} \\
      \hline
      $E(\mathbf{a},\mathbf{b},\mathbf{c})$ & $E(\mathbf{a},\mathbf{b'},\mathbf{c'})$ & $E(\mathbf{a'},\mathbf{b},\mathbf{c'})$ & $E(\mathbf{a'},\mathbf{b'},\mathbf{c})$ \\
      \hline
      $0.667 \pm 0.079$ & $-0.568 \pm 0.096$ & $-0.614 \pm 0.084$ & $-0.674 \pm 0.075$ \\
      \hline
      \multicolumn{4}{|c|}{$M$} \\
      \hline
      \multicolumn{4}{|c|}{$2.521 \pm 0.167$} \\
      \hline
      \multicolumn{4}{|c|}{\textbf{Experiment II (deterministic)}} \\
      \hline
      $E(\mathbf{a},\mathbf{b},\mathbf{c})$ & $E(\mathbf{a},\mathbf{b'},\mathbf{c'})$ & $E(\mathbf{a'},\mathbf{b},\mathbf{c'})$ & $E(\mathbf{a'},\mathbf{b'},\mathbf{c})$ \\
      \hline
      $0.526 \pm 0.138$ & $-0.765 \pm 0.111$ & $-0.514 \pm 0.141$ & $-0.786 \pm 0.117$ \\
      \hline
      \multicolumn{4}{|c|}{$M$} \\
      \hline
      \multicolumn{4}{|c|}{$2.590 \pm 0.255$} \\
      \hline
  \end{tabular}
  \caption{Measured Mermin parameters using random and deterministic basis choices.}
  \label{tab.DeterministicBasisChoicesExpectationsAndMermin}
\end{table}

%===============================================================================
% Bibliography
%===============================================================================

\bibliography{GHZ_locality}
\bibliographystyle{naturemag}
% The bibliography output from the GHZ_locality.bbl file

%===============================================================================
% Acknowledgements
%===============================================================================

\begin{acknowledgments}
\textbf{Acknowledgements:} The authors thank Deny Hamel and Thorald Bergmann for technical discussions; John Dengis and Catherine Holloway for assistance in the laboratory; Mike Ditty, Feridun Hamdullahpur, Dennis Huber, Dan Parent, Rick Zalagenas, Mike Lazaridis, and George Dixon for their support in gaining access to the roof of RAC; Peter Fulcher for safety training; Adrian Conrad, AGFA, and Carol Stewart for allowing access to private property; Bob Zinger, Gerry Doyle, Dave Copeland, Rick Reger, Bruce Mill, Tom Galloway, and Zhenwen Wang for carpentry and electronics; Steve Payne at Leysop Ltd for assistance with Pockels cells; Mark Morelli at UBNT.ca for technical advice on wireless networks; Roncare and UW plant operations for snow plowing, and good samaritan Matt Seibel for towing. The authors are grateful for financial support from NSERC, CRC, CFI, Industry Canada, CIFAR, OCE, and QuantumWorks. RP acknowledges support from the FWF (J2960-N20), MRI, the VIPS Program of the Austrian Federal Ministry of Science and Research and the City of Vienna as well as the European Commission (Marie Curie, FP7-PEOPLE-2011-IIF).

\textbf{Author Contributions:} CE, RL, GW, TJ, and KJR conceived the experiment; CE, EMS, KF, JL, BLH, ZY, CP, JPB, RP, LR, and NG constructed the experiment; BLH and LKS performed the space-time analysis; CE, EMS, KF, JL, CP, and JPB took the data; CE analyzed the data; All authors contributed to the manuscript.
\end{acknowledgments}

%===============================================================================
% Supplementary Info
%===============================================================================

\end{document}